\documentclass[
 preprint,
 amsmath,amssymb,
 aps,
 prx,
]{revtex4-2}

\usepackage[pagewise]{lineno}

\usepackage[utf8]{inputenc}
\usepackage[T1]{fontenc}
\usepackage{geometry}
\usepackage{graphicx}
\usepackage{amsmath, amssymb}
\usepackage{hyperref}

\begin{document}

\title{Self-propulsive active nematics}

\author{Niels de Graaf Sousa}
\affiliation{Niels Bohr Institute, University of Copenhagen, Copenhagen, Denmark}
\author{Simon Guldager Andersen}
\affiliation{Niels Bohr Institute, University of Copenhagen, Copenhagen, Denmark}
\author{Aleksandra Ardaševa}
\affiliation{Niels Bohr Institute, University of Copenhagen, Copenhagen, Denmark}
\author{Amin Doostmohammadi}
\email[]{doostmohammadi@nbi.ku.dk}
\affiliation{Niels Bohr Institute, University of Copenhagen, Copenhagen, Denmark}

\date{\today}

\begin{abstract}
Increasing evidence suggests that active matter exhibits instances of mixed symmetry that cannot be fully described by either polar or nematic formalism. Here, we introduce a minimal model that integrates self-propulsion into the active nematic framework. Our linear stability analyses reveal how self-propulsion shifts the onset of instability, fundamentally altering the dynamical landscape. Numerical simulations confirm these predictions, showing that self-propulsion induces anti-hyperuniform fluctuations, anomalous long-range order in vorticity, and non-universal self-similar energy cascades. Notably, these long-range ordered states emerge within the active turbulence regime well before the transition to a flocking state. Additionally, our analyses highlight a non-monotonic dependence of self-organization on self-propulsion, with optimal states characterized by a peak in correlation length. These findings are relevant for understanding of active nematic systems that self-propel, such as migrating cell layers or swarming bacteria, and offer new avenues for designing synthetic systems with tailored collective behaviours, bridging the gap between active nematics and self-propulsive systems.
\end{abstract}

\maketitle

\section{\label{sec: intro}Introduction}

Active matter, encompassing systems driven far from equilibrium by internal energy dissipation, has transformed our understanding of collective dynamics in both biological and synthetic contexts \cite{needleman2017active}. From bacterial colonies \cite{Igor_Active_Bacterial_Review,PNAS_living_liquid_crystals} and cellular assemblies \cite{Blanch_Mercader_2018,cell_henkes_2020}, to synthetic microswimmers \cite{gompper20202020,Z_ttl_2016}, active matter exhibits a remarkable variety of behaviours that challenge traditional frameworks of condensed matter physics. These systems provide a fertile arena for exploring new phases, instabilities, and emergent phenomena, enabling insights into both natural and engineered processes. Among these, active nematics -- materials composed of elongated, interacting active particles -- stand out for their ability to describe a wide range of experimental systems, from synthetic liquid crystals to biological tissues \cite{saw2018biological, doostmohammadi2022physics, yaman2019emergence}. Active nematics capture hallmark features of active flows, such as defect dynamics \cite{Giomi_2014}, spontaneous turbulence \cite{Alert_2022}, and energy transfer across scales \cite{ActiveNematicsADJIJYFS,pearce2024topologicaldefectsleadenergy}, making them a cornerstone in the study of active matter.

Despite their success, the active nematic framework has an inherent limitation. While it accounts for self-propulsion through active stresses, it fails to capture a crucial aspect of self-propulsion: the ability of active particles to spontaneously move along the direction of their polarity. This capability, essential to motile biological entities such as bacteria \cite{meacock2021bacteria, copenhagen2021topological} and motile eukaryotic cells \cite{trepat2018mesoscale, kawaguchi2017topological}, e.g. sperm cells \cite{Turbulence_of_swarming_sperm}, introduces unique dynamics and instabilities that remain unexplored in standard active nematic formulations. These examples demonstrate that, unlike existing theoretical models which are bound by a specific individual symmetry, experimental realizations can exhibit mixed symmetry, where the nematic symmetry associated with the orientation of active particles inevitably co-exists with the polar symmetry of the particles motility.

Few theoretical studies have addressed the interplay between nematic symmetry and self-propulsion. Among these, models of self-propelled hard rods \cite{baskaran2008hydrodynamics,baskaran2012self,Self_propelled_rods_review_Bår,vafa2025phase} use a set of coupled equations to describe the polarity vector and nematic tensor fields. Another notable approach, based on the polar formalism, incorporates a Frank elastic term to account for nematic inhomogeneities in the order field \cite{amiri2022unifying}. This model successfully predicts the emergence of both  half- and full-integer topological defects. However, the absence of such self-propulsive motion in current active nematic models restricts their applicability to systems where translational self-propulsion and nematic order coexist. 

Here, we introduce a simple model that integrates self-propulsion into the active nematic framework. Through linear stability analyses, we show that accounting for self-propulsion alters the instability threshold, reshaping the dynamical landscape of active nematics. Our numerical simulations validate these theoretical predictions, revealing additionally that self-propulsion triggers anti-hyperuniform fluctuations of topological defects, enhances nematic order and leads to anomalous long-range vorticity correlations, as well as non-universal self-similar energy cascades. These ordered states emerge within the active turbulence phase, preceding the transition to a flocking state. Our findings underscore the intricate relationship between self-organization and self-propulsion, with optimal states marked by a peak in correlation length.

\section{\label{sec: model} Methods}
The model presented here builds on the well-established continuum representation of active nematics \cite{ActiveNematicsADJIJYFS}. Particle orientation is described by the nematic director, $\hat{n}$, which takes into account the nematic symmetry $\hat{n}=-\hat{n}$. The nematic tensor is defined as $Q_{ij}=2 q (n_i n_j -\frac{\delta_{ij}}{2})$ with $q$ being the magnitude of order. The nematic order parameter evolves according to the Beris--Edwards equation:
\begin{equation}
    \label{eq: dynamical equation of nematic tensor}
    \partial_t Q_{ij}+(v_k+V_0 p_k)\partial_k Q_{ij}=\Gamma H_{ij}+S_{ij}. 
\end{equation}
Here, $S_{ij}$ is the co-rotational term, which has the following form:
\begin{equation}
    \label{eq: corrotational term}
    \begin{split}
         S_{ij}=(\lambda E_{ik}+ \Omega_{ik})(Q_{kj}+\frac{\delta_{kj}}{2})
         +(Q_{ik}+\frac{\delta_{ik}}{2})(\lambda E_{kj} -\Omega_{kj})
    -2\lambda(Q_{ij}+\frac{\delta_{ij}}{2})(Q_{kl}\partial_k v_l),
    \end{split}  
\end{equation}
with $E_{ij}=\frac{1}{2}(\partial_i v_j + \partial_j v_i)$ and $\Omega_{ij}=\frac{1}{2}(\partial_j v_i - \partial_i v_j)$ being the strain rate and vorticity tensors, respectively. $H_{ij}=-\left(\frac{\delta \mathcal{F}}{\delta Q_{ij}}\right)$ is the molecular field that accounts for the free energy, $\mathcal{F}$, relaxation. The free energy contains the Landau--de Gennes expansion with coefficient, $A$, and the Frank elastic term with constant, $K$:
\begin{equation}
    \label{eq: free energy}
    \mathcal{F}=\int dA\left[\frac{K}{2}(\partial_k Q_{ij})^2+A(1-Q_{ij}Q_{ji})^2\right].
\end{equation}

\begin{figure}[htb!]  
\centering
    \includegraphics[width=0.5\linewidth]{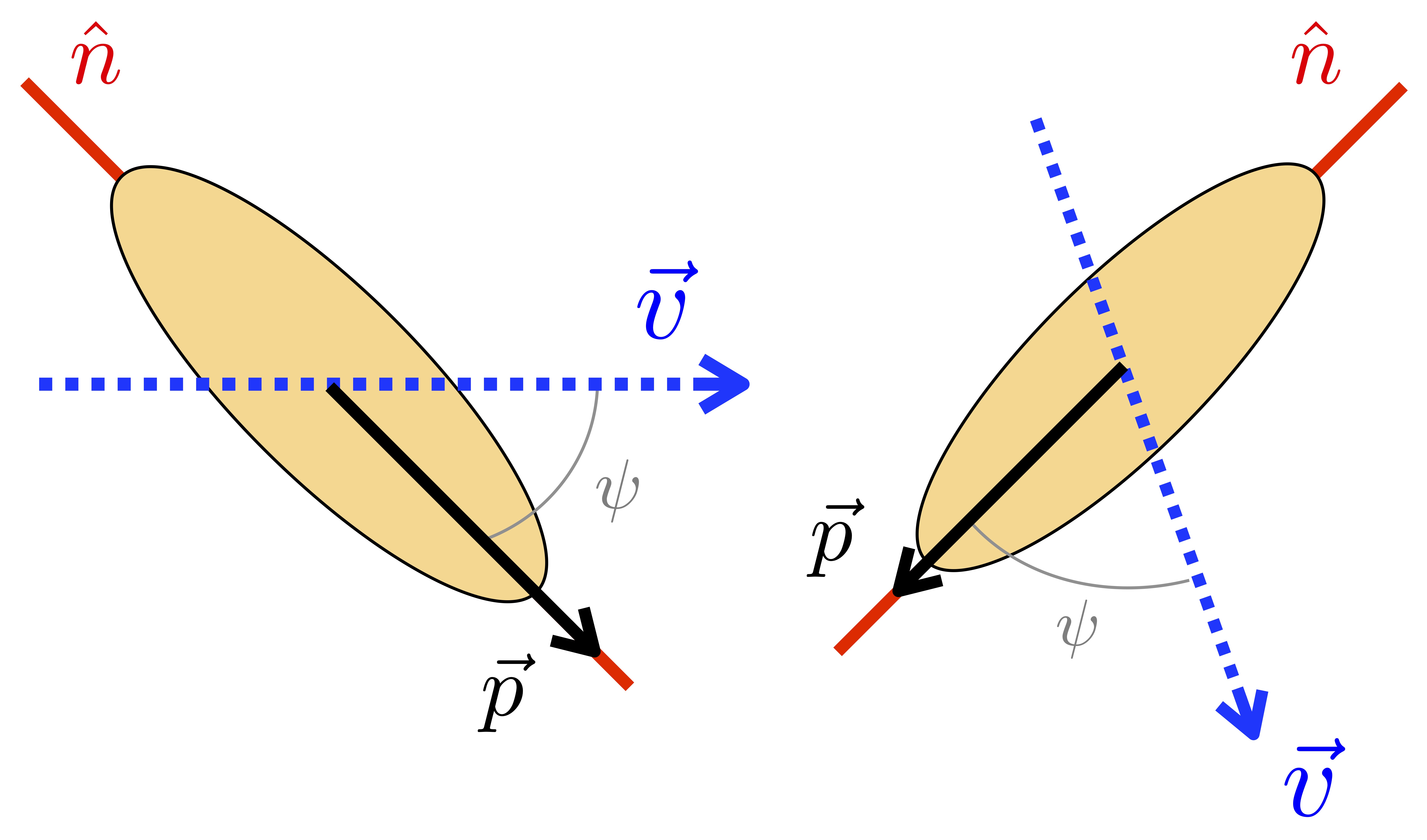}
    \caption{\textbf{Polarity assignment mechanism}. The dotted blue arrow represents the flow velocity field, $\Vec{v}$, the red stripe the nematic director, $\hat{n}$, and the black arrow shows the assigned polarity, $\Vec{p}$. Polarity is assigned at each time step by determining which direction of $\hat{n}$ is closest to the flow field, indicated by the angle $\psi$.}
\label{fig: Schematic polarity assignemnt}
\end{figure}

Additionally, we introduce in Eq. \ref{eq: dynamical equation of nematic tensor} a new self-advective term, $V_0 p_k$, which breaks the nematic symmetry. This term represents the self-propulsion of the particles migrating at speed $V_0$, referred to as the self-propulsion speed. This is a common form of implementing self-propulsion in polar systems \cite{Giomi_Polar_LSA,marchetti2013hydrodynamics}, and describes self-propulsion along the polarity vector through a self-advective term. As such we expect this form of implementing self-propulsion to be applicable to systems with different mechanisms of motility generation. We take advantage of the fact that rod-shaped particles can polarize in two distinct directions and, therefore, for each active particle there are only two options for the polarity direction. In this study, we assume that active particles polarize in the direction that has the least deviation from the fluid flow. Thus, polarity is assigned to the direction of $\hat{n}$ closest to the fluid flow, as illustrated in the schematic in Fig. \ref{fig: Schematic polarity assignemnt}. The assumption is based on experimental realizations that report active particles align their polarity with the fluid flow. Some examples include eukaryotic cells, such as \textit{Dictyostelium discoideum}, \cite{Decave2003} and epithelial monolayers, which have been observed to align the direction of their lamellipodia and consequently their intrinsic polarity with the total force acting on the cells \cite{peyret2019sustained}. This behaviour is also shared for bacterium, such as \textit{Myxococcus xanthus}, where each individual particle aligns with the flow direction \cite{han2023local}. 

The fluid velocity, $v_i$, evolves according to the incompressible Navier-Stokes equations:
\begin{align}
    \label{eq: Navier-Stokes}
    \rho(\partial_t v_i + v_k\partial_k v_i)&= \partial_j \sigma_{ij}, \\
    \partial_i v_i&=0,
\end{align}
where $\rho$ is the density and $\sigma_{ij}$ is the stress tensor, which has three contributions -- viscous, elastic, and active:
\begin{equation}
    \label{eq: nematic stress tensor}
    \begin{aligned}
        \sigma^{\text{viscous}}_{ij} &= 2\eta E_{ij}, \\
        \sigma^{\text{passive}}_{ij} &= -P \delta_{ij} + 2 \lambda \left( Q_{ij} + \frac{\delta_{ij}}{2} \right) Q_{lk} H_{kl} 
         - \lambda H_{ik} \left( Q_{kj} + \frac{\delta_{kj}}{2} \right) - \lambda \left( Q_{ik} + \frac{\delta_{ik}}{2} \right) H_{kj} \\
        &\quad - \partial_i Q_{kl} \frac{\delta \mathcal{F}}{\delta \partial_j Q_{lk}} + Q_{ik} H_{kj} - H_{ik} Q_{kj}, \\
        \sigma^{\text{active}}_{ij} &= -\zeta Q_{ij},
    \end{aligned}
\end{equation}
 with $\zeta$ being the dipolar activity coefficient, which describes the amount of ambient free energy. $\lambda$ is the flow alignment coefficient and $\eta$ is the fluid viscosity.

The Navier--Stokes equations (Eq. \ref{eq: Navier-Stokes}) are solved using the lattice Boltzmann method \cite{LatticeBoltzmann} and the advection--diffusion equation for $Q_{ij}$ (Eq. \eqref{eq: dynamical equation of nematic tensor}) is solved using finite--difference method. The equations are solved numerically over a square lattice of length, $L$, with periodic boundary conditions. The simulation variables are the following unless stated otherwise: $\rho=40$, $\eta=20$, $\lambda=0.2$, $\Gamma=0.1$, $K=0.05$, $A=0.1$, $L=1024$ and $\zeta=0.05$. For the Linear Stability section \ref{sec: LSA}, the equations have been implemented in OpenFOAM-v2306.

\section{\label{sec: /Results}Results}
\subsection{\label{sec: LSA}Linear Stability Analysis}
To shed light on the impact of including self-propulsion effect in active nematics formulation, we perform a linear stability analysis of the nematic state for the modified equations. The classical nematic instability arises from the interplay between active stress and shear flow \cite{Giomi_LSA_Nematic,Nematic_LSA_Iso}. As we shall show, accounting for self-propulsion drastically affects this instability. The dynamical variables that drive the instability are the off-diagonal nematic tensor component, $Q_{xy}$, and the vorticity, $\omega$, strategically defined as $\omega=\partial_x v_y -\partial_y v_x$. The diagonal terms of the nematic tensor, $Q_{xx}$, are disregarded since the transverse and longitudinal modes decouple from the Jacobian matrix, as previously shown by \cite{Giomi_LSA_Nematic}. 

We initialise the systems as a homogeneously ordered state along the x-axis. At zeroth order, the nematic director has the following components, $n_x=1$ and $n_y=0$, and the system has no net vorticity, i.e. $\phi^0=(Q_{xx}, Q_{xy},\omega)=(1,0,0)$. We apply an infinitesimal perturbation to this steady-state system, which accounts for the periodic boundary conditions on the square lattice:
\begin{equation}
    \label{eq: infinitesimal pertorbation LSA}
    \delta \phi^1=\phi_{nm}(t)e^{iq(nx+my)},
\end{equation}
with $q=\frac{2 \pi}{L}$ being the shortest wave vector in the system and $L$ -- the longitude of the system. Therefore, our set of dynamical variables with the added perturbation has the following form: $\phi(x,t)= \phi^0+\delta \phi^1=(1+\delta Q_{xx},\delta Q_{xy}, \delta \omega)$. As the system starts from an $x$-aligned state, the polarization takes the form $(p_x,p_y)=(1,0)$, where we assume that the flow field has a positive $x$-component. 

Linearizing the hydrodynamic equations reduces them to a set of coupled linear ordinary differential equations, $\partial_t \phi_{nm}=\mathcal{J}_{nm} \phi_{nm}$ (see Appendix \ref{sec: LSA appendix} for details). For the instability to rise, the real part of the eigenvalues of $\mathcal{J}_{nm}$ has to be positive. The first mode to rise is the longitudinal $(n,m)=(1,0)$ with the following eigenvalue expression:
\begin{equation}\label{eq: eigenvalue long}
\begin{split}
    \Lambda_{10}=&\frac{1}{2}q \biggl[ -q\left(\Gamma K +\frac{\eta}{\rho}\right)-i V_0 \\
    &\pm \sqrt{q^2\left(\Gamma K - \frac{\eta}{\rho}\right)^2+2q V_0\left(\Gamma K-\frac{\eta}{\rho}\right) i-V_0^2 +\frac{2(2+\lambda)}{\rho}[\zeta-q^2 K (2+\lambda)]}\; \biggr].
\end{split}
\end{equation}

\begin{figure}[htb!]  
\centering
    \includegraphics[width=0.6\linewidth]{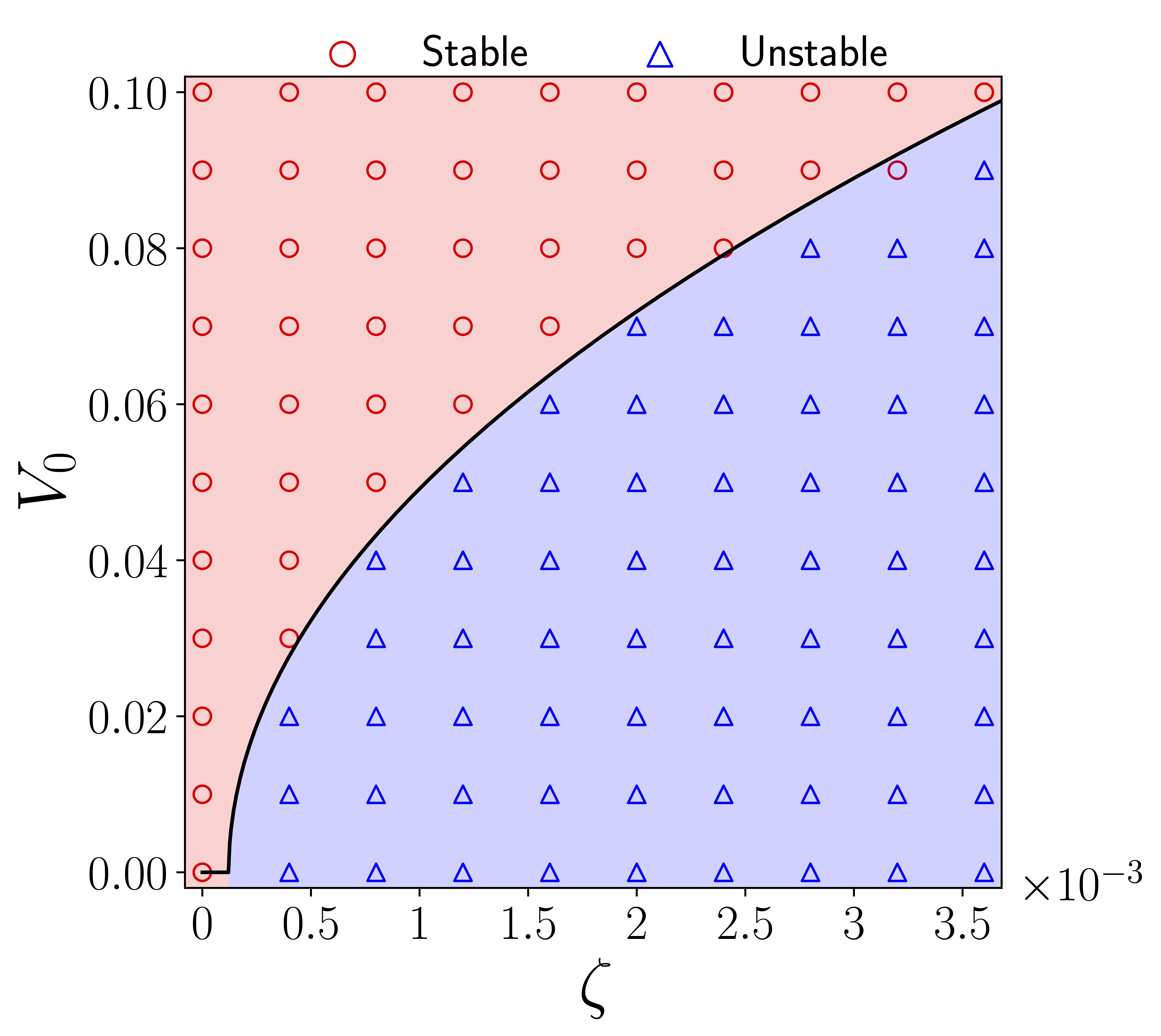}
    \caption{\textbf{Stability of the $(V_0,\zeta)$ phase space}. The black line is the critical value of self-propulsion, $V_0^{*}$, needed to raise the instability Eq. \ref{eq: critical value of alpha}. Scatter points are obtained through simulations: red circles indicate that the system is stable and blue triangles that the system is unstable.}
\label{fig: LSA phase space}
\end{figure}

The values of $V_0$ that suppress the instability are given by the condition $Re[\Lambda_{10}]<0$. As a result, one can find that the instability is suppressed above certain values of $V_0^{*}$:
\begin{equation}
    \label{eq: critical value of alpha} 
    V_0^{*}=\left(\Gamma K +\frac{\eta}{\rho}\right)\sqrt{\frac{(2+\lambda)}{2 \Gamma K \eta}[\zeta-q^2K (2+\lambda)]-q^2}.
\end{equation}
This expression divides the phase space into stable and unstable regions (\textit{black line} in Fig. \ref{fig: LSA phase space}) and shows close agreement with the results obtained from numerical simulations. This result proves that $V_0$ can suppress the instability and provides a new pathway on how systems can become unstable. This aligns with the results of the pure polar formalism, where self-propulsion has been shown to suppress the polar instability \cite{Giomi_Polar_LSA,LSA_Polar_Flocking_Simha_R}. 

Introducing self-propulsion allows the definition of a characteristic time scale for the nematic tensor, given by $\tau_Q= \Gamma K / V_0^2$. This can be compared to the other characteristic time scale set through the balance of active and viscous stress, $\tau_v= \eta /\zeta$. When the time scales are equal, it establishes a connection between the self-propulsion speed and the activity coefficient, $V_0 \propto \sqrt{\zeta}$. Notably, this dependency is also followed for the critical value of $V_0^*$, which suppresses the instability in Eq. \ref{eq: critical value of alpha}, setting a scaling relation between both variables.

\subsection{Self-propulsion enhances order}
\begin{figure}[htb!]  
\centering
    \includegraphics[width=1\linewidth]{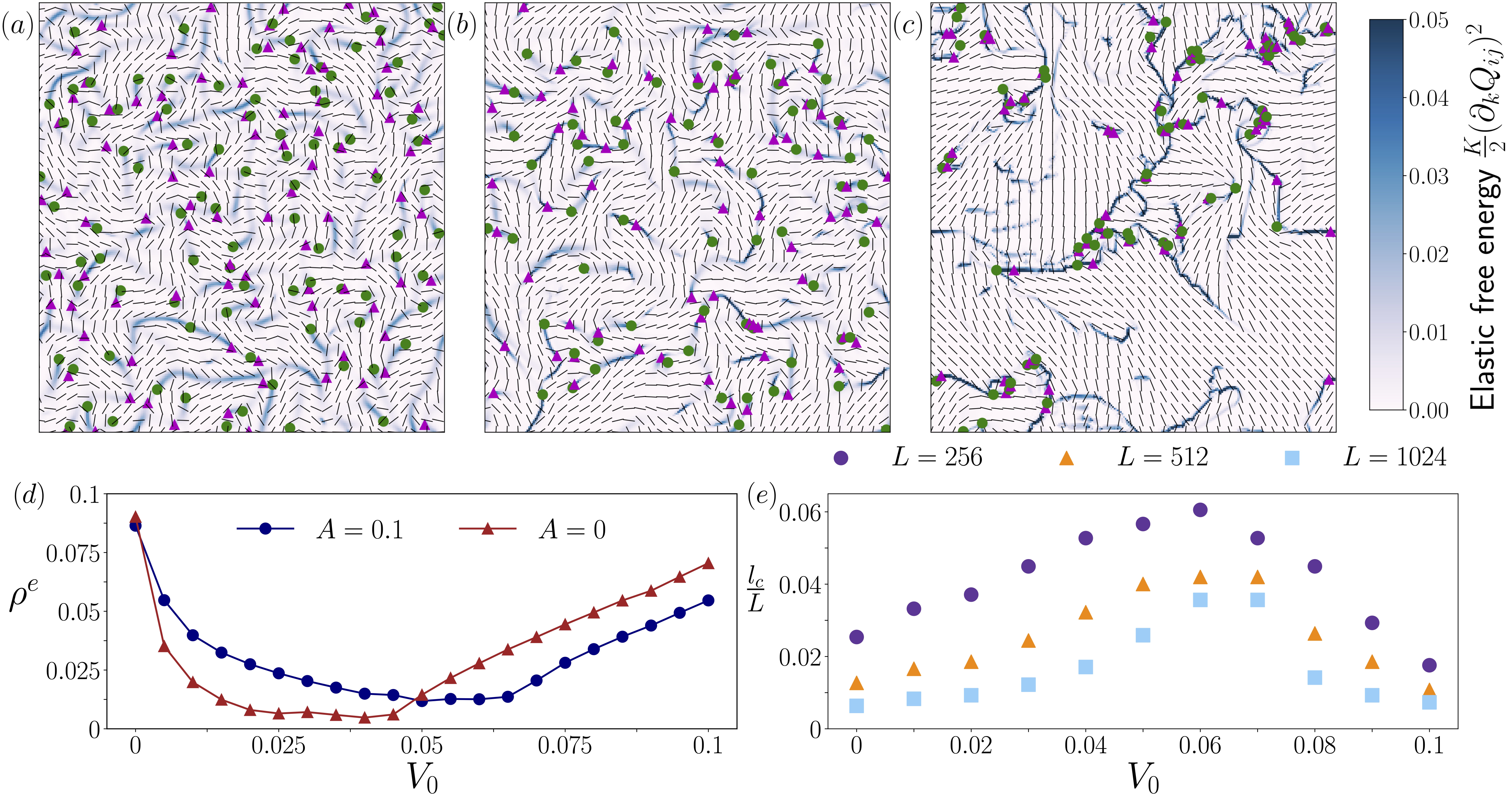}
    \caption{\textbf{Self-propulsion enhances order}. Snapshots of simulations with increasing values of self-propulsion: (a) $V_0=0$, (b) $V_0=0.01$, and (c) $V_0=0.06$. Black lines denote the nematic director, green circles and pink triangles represent $+1/2$ and $-1/2$ topological defects, respectively. The colour denotes the magnitude of the elastic free energy, $E=\frac{K}{2}(\partial_k Q_{ij})^2$. (d) The elastic free energy density, $\rho^e$, as a function of self-propulsion speed, $V_0$, for two different bulk free energy coefficients, $A=0.1$ and $A=0$. (e) Correlation length $l_c$ as a function of self-propulsion speed $V_0$ for different system sizes $L$. $l_c$ is defined as the distance at which the correlation function decays by a factor $e^{-1}$. (d--e) averaged over the lattice and different conformations.}
    \label{fig: OrderEnhancment}
\end{figure}

To move beyond the linear analyses and demonstrate the effect self-propulsion has on the collective motion of active nematic, we numerically investigate the emergent dynamics. First, we consider different values of self-propulsion, $V_0$. As shown in Fig. \ref{fig: OrderEnhancment}a--c, increasing $V_0$ leads to the enhancement of nematic order. In particular, the areas of nematic alignment grow and the defects become more closely spaced while being confined to the boundaries between the nematic regions. Significantly, this effect arises at the chaotic flow regime, well before the flocking transition, with the dipolar activity coefficient set to $\zeta=0.05$, an order of magnitude higher than the values shown in Fig. \ref{fig: LSA phase space}. To quantify the enhancement of the nematic order, we define the elastic free energy density, $\rho^e= \left\langle \frac{E}{max(E)} \right\rangle$, with $E=\frac{K}{2}(\partial_k Q_{ij})^2$ being the elastic free energy and compute the averages over different conformations and the whole lattice (Fig. \ref{fig: OrderEnhancment}d). The function exhibits a convex form with the minimum occurring at $V_0 \approx 0.06$ and increasing thereafter. This indicates that the self-propulsion speed of the particles can regulate the system’s order. Particularly, this order enhancement and following decay, driven by  $V_0$, can occur in the absence of bulk coefficients, $A=0$ (\textit{red line} Fig. \ref{fig: OrderEnhancment}d). Notably, this finding is independent of the order induced by active stress \cite{Thampi_2015_Intrinsinc_Free_Energy}, demonstrating that self-propulsion alone can modulate the systems' order. Strikingly, unlike active stress, self-propulsion exhibits a non-monotonic behaviour, indicating that increasing $V_0$ can both enhance and break nematic order.

To further characterize the emergent nematic order, we calculate the correlation length, $l_c$, defined as the distance at which the autocorrelation of the nematic director, $C_{nn}(r)$, decays by a factor of $e^{-1}$, these measurements provide information on the average length scale over which the nematic director remains aligned. Fig. \ref{fig: OrderEnhancment}e presents the correlation length as a function of the self-propulsion coefficient for various system sizes, $L$. It is evident that as the self-propulsion coefficient $V_0$, increases, the correlation length also increases, reaching a maximum at $V_0 \approx 0.06$ -- the same turning point observed in the free energy density (Fig. \ref{fig: OrderEnhancment}d). Beyond this point, the correlation length decreases, exhibiting a concave behaviour. This indicates that self-propulsion can both establish and disrupt long-range order depending on its magnitude, irrespective of the system size. Although the correlation lengths on both sides of the maxima are similar, the system dynamics differs significantly, as is detailed in the following sections.

\subsection{Defect anti-hyperuniformity}
As stated in the previous section and in Fig. \ref{fig: OrderEnhancment}a--c, the enhancement of nematic order is regulated by the self-propulsion speed. In this section, we show that this non-monotonic behaviour is rooted in the defect configuration, as the defects become more closely spaced while being restricted to the boundaries of the nematic regions. Such spatial irregularity is inherently associated with hyperfluctuations and anti-hyperuniformity \cite{torquato2018hyperuniform, marchetti2013hydrodynamics}, motivating us to examine the spatial behaviour of defect density fluctuations at long range. 

Based on this scaling, systems of particles can be classified according to their asymptotic density fluctuation behaviour as either \textit{uniform}, \textit{hyperuniform} or \textit{anti-hyperuniform}. Most disordered states of matter, e.g. ordinary fluids and amorphous solids, are uniform, meaning that $\sigma_N^2(R)$, the variance of the number of particles contained in a randomly placed spherical observation window with radius, $R$, scales like $\sigma_N^2(R) \sim R^{d-\alpha}$, where $\alpha = 0$, in the limit of large $R$. For disordered hyperuniform systems, $1 > \alpha > 0$, whereas $-d \leq \alpha < 0$ for anti-hyperuniform systems \cite{torquato2021numberfluctuations}. While equilibrium examples of anti-hyperuniformity are limited to systems at thermal critical points \cite{torquato2018hyperuniform}, many active systems exhibit this property as a manifestation of ``giant number fluctuations" \cite{marchetti2013hydrodynamics}. In this context, the relation $\sigma_N^2(R) \sim R^{d-\alpha}
$ is recast as $\sqrt{\sigma_N^2(R)} = \langle N(R) \rangle ^\beta$, where $\beta = 0.5(1 - \alpha/d)$  \cite{toner1995long-range}. For uniform systems, $\beta = 0.5$, as predicted by the law of large numbers, whereas $\alpha < 0$ corresponds to $\beta > 0.5$, with the implication that the error on the sample mean improves more slowly than $1/\sqrt{N_{\textup{samples}}}$. The scaling behaviour of density fluctuations can equivalently be described through the asymptotic behaviour of the static structure factor, $S$. In particular, for disordered hyperuniform and anti-hyperuniform systems, $S(\mathbf{q}) \sim |\mathbf{q}|^{\alpha}$ and $\sigma_N^2(R) \sim R^{d-\alpha}$ are equivalent in the infinite volume limit \cite{torquato2018hyperuniform}.

\begin{figure}[htb!]  
\includegraphics[width=\linewidth]{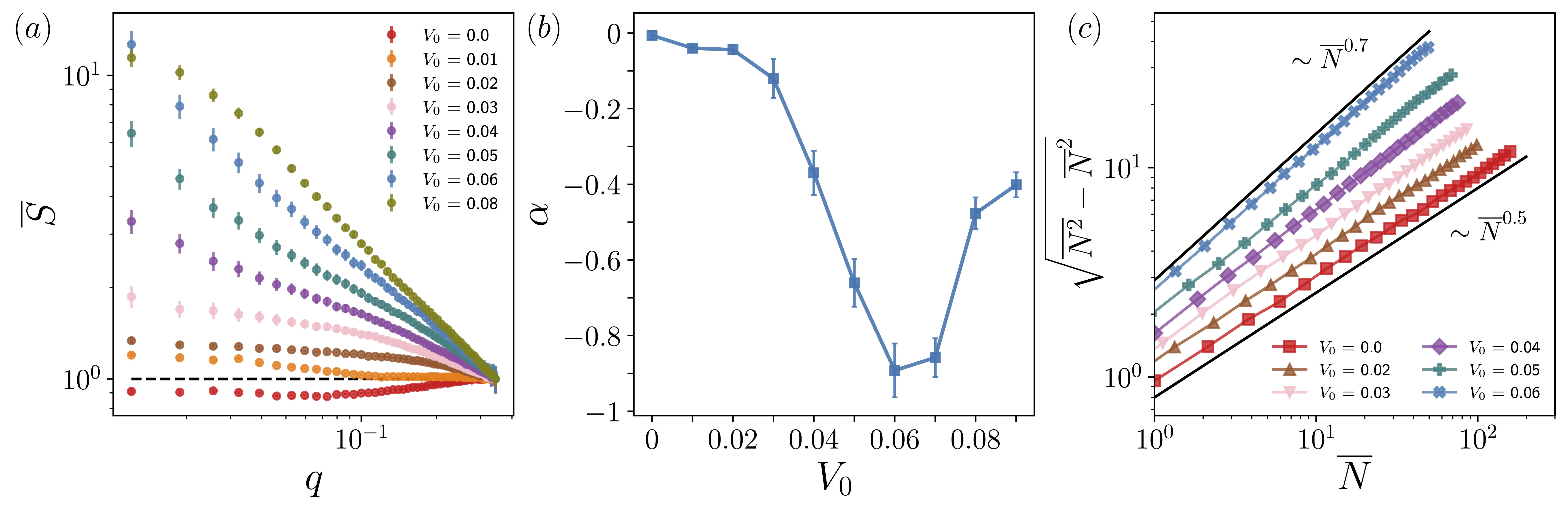}
\caption{\textbf{Structure factor, scaling exponents, and giant number fluctuations of topological defects for $L=1024$.} (a) Normalized structure factor, $\overline{S}$, (averaged over orientations and time) as a function of the wave vector norm, $q$, across $V_0$. The dashed line at $S = 1$ indicates the theoretical value of the structure factor for the Poisson process. (b) Estimated scaling exponents, ${\alpha}$, as obtained by fitting $\overline{S}(q) \sim q^{\alpha}$ for small $q$.  (c) Square root of the number variance, $(\overline{N(R)^2} - \overline{N(R)}^2)^{1/2}$, as a function of the average number of points, $\overline{N(R)}$, within a randomly placed spherical window with radius, $R$, for several radii, $R$. Giant number fluctuations with $\beta \approx 0.7 \Rightarrow \alpha \approx -0.8$ are observed for $V_0 = 0.06$, whereas for $V_0 = 0$, the expected $\beta = 1/2$ scaling of uniform systems is observed. }
\label{fig: Anti-hyperuniformity}
\end{figure}

For every self-propulsion speed and time step, we estimate the structure factor of the corresponding defect configuration, $S(\mathbf{q})$, for all wave vectors that satisfy $|\mathbf{q}| \leq q_{\text{upper}} = 1/4$ (see Appendix \ref{sec: sfac appendix} for details). These estimates are then averaged over time and orientations to obtain $\overline{S}(q)$ (Fig. \ref{fig: Anti-hyperuniformity}a). The dashed horizontal line at $S=1$ indicates the expected behaviour for the Poisson point process, and we see that for $V_0 \leq 0.02$, $(q\rightarrow0)S(q)\propto q^{\alpha}$ with $\alpha \approx 0$. For $V_0 > 0.02$, $\alpha < 0$ and the value of $\overline{S}(q)$ generally increases with $V_0$.  For $V_0 \geq 0.08$, however, we note that the slope of $\overline{S}$  flattens for small $q$, indicating that $\overline{S}$ might not diverge as $q \rightarrow 0$, i.e. that hyperfluctuations do not persist in the infinite-volume limit. In contrast, for $V_0 \in [0.4, 0.6]$, the slope of $\overline{S}$ steepens for small $q$, suggesting that hyperfluctuations persist at long range and so that defects are in fact anti-hyperuniform for these values of $V_0$.

\noindent
This difference in behaviour is reflected in the corresponding scaling exponents, $\alpha$, which are obtained by fitting $S(q) \sim q^{\alpha}$ for small $q$ (Fig. \ref{fig: Anti-hyperuniformity}b). We see that anti-hyperuniformity is most pronounced at $V_0=0.06$, the self-propulsion speed of maximal order, and that the strength of anti-hyperuniformity increases as $V_0 \rightarrow 0.06$ from either side.

To directly measure defect density fluctuations, we estimate the moments of $N(R_i)$, the number of points within a randomly placed spherical observation window with radius, $R_i$, for 25 radii $R_i \in [L/100, L/10]$ (see Appendix \ref{sec: number variance appendix} for details).  As expected, the scaling of $\left(\sigma_N^2(R)\right)^{1/2}$ reveals that giant number fluctuations are indeed present at $V_0 = 0.06$, with $\beta(V_0=0.06) \approx 0.7 \Rightarrow \alpha(V_0=0.06) \approx -0.8$ (Fig. \ref{fig: Anti-hyperuniformity}c), in agreement with the structure factor estimate, $\alpha(V_0=0.06) = - 0.89 \pm 0.07$. Similarly, for $V_0=0$ we observe the ${N}^{1/2}$ scaling characteristic of uniform systems.

Taken together, these findings establish that defect configurations exhibit anti-hyperuniformity and giant number fluctuations for self-propulsion speeds near $V_0=0.06$. 
Importantly, the observation that for $V_0 \in [0.04,0.06]$, the slope of $\overline{S}$ steepens as $q$ is decreased suggesting that hyperfluctuations persist at long range, i.e. that
these anti-hyperuniform states are intrinsic to the system and not merely finite-size effects. Strikingly, the onset and strength of anti-hyperuniformity mirror those of long-range order (Figs. \ref{fig: OrderEnhancment}e \& \ref{fig: Flow}b). This underscores the critical role of the self-propulsion speed, $V_0$, in governing not only nematic order but also the spatial organization and fluctuation dynamics of topological defects.

\subsection{Flow dynamics}
Having described the effect self-propulsion has on nematic order and defect space configuration, we now consider the impact it has on the flow dynamics.

\begin{figure}[htb!]
    \includegraphics[width=1\linewidth]{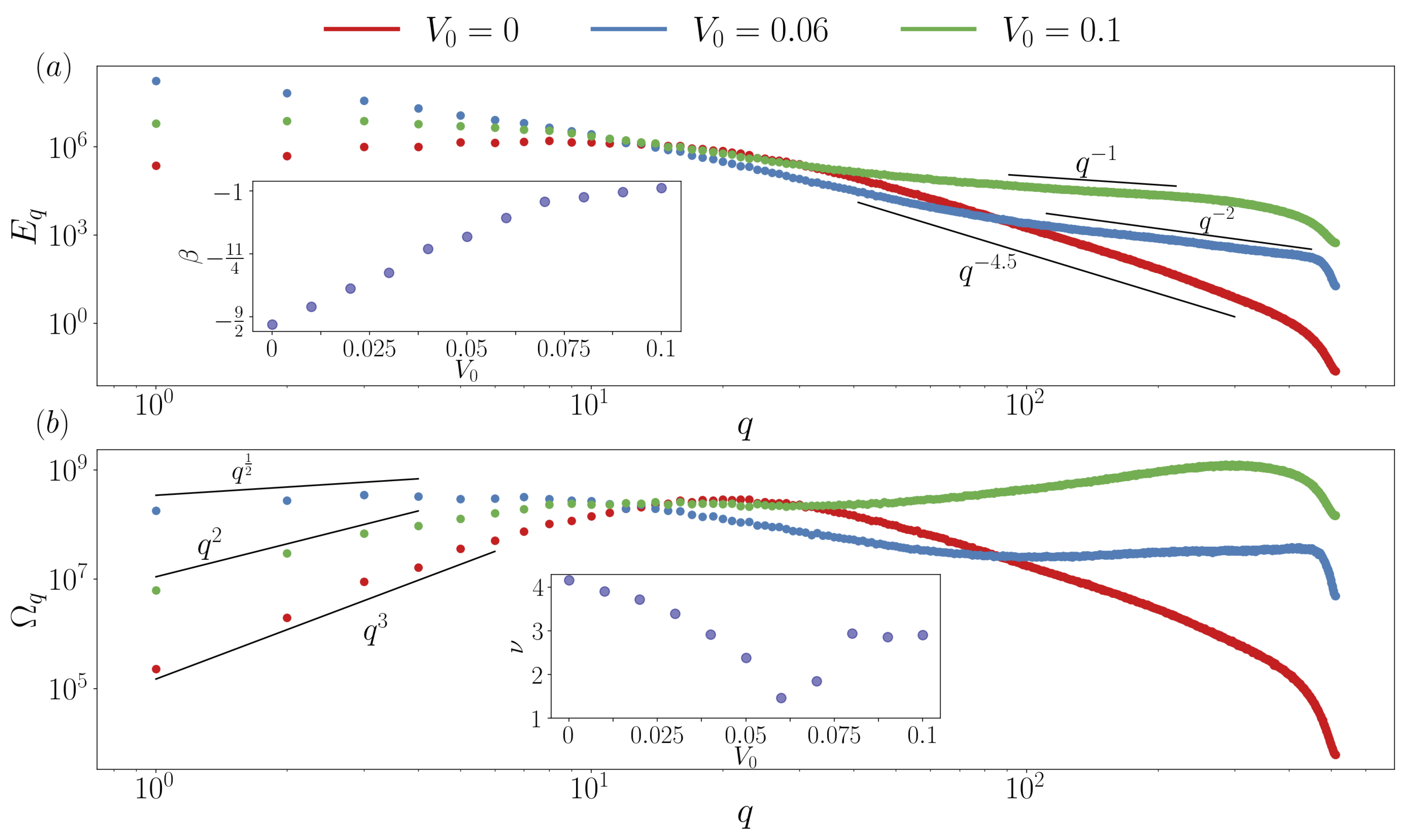}
    \caption{\textbf{Non-universal energy cascades and anomalous long-range order}. (a) Kinetic energy spectra as a function of the wave vector, $q$, for different self-propulsion speeds, $V_0$. Inset plot shows the scaling exponents, $\beta$, for large wave vectors as a function of $V_0$, $E(q) \propto q^{\beta}$. (b) Enstrophy energy spectra as a function of the wave vector, $q$, for different self-propulsion speeds, $V_0$. Inset plot shows the scaling exponents, $\nu$, for large distances as a function of $V_0$, $\langle \omega^2(r) \rangle \propto r^{-\nu}$. Wave vector has been normalized by the smallest wave vector $q=\frac{2 \pi}{L}$.}
\label{fig: Flow}
\end{figure}

First, we perform a spectral analysis to display how the energy transfer varies from the well-known nematic energy cascade due to the self-propulsion speed. Figs. \ref{fig: Flow}a \& \ref{fig: Flow}b show the kinetic energy spectra, $E_q \propto \int \langle \lvert v(q) \rvert^2 \rangle d \Vec{q}$,  for various self-propulsion speeds. For $V_0=0$, the well-established scalings are retrieved \cite{giomi_2015_defect_statistics}. The inset plot in Fig. \ref{fig: Flow}a illustrates the scaling, $E(q) \propto q^{\beta}$, for small lengths ($q>1$) as a function of $V_0$ demonstrating how the scaling grows linearly with self-propulsion due to the energy injection that augments the advection of the nematic tensor, describing a non-universal turbulence scaling law \cite{EFrey_New_Class_turbulence}. Furthermore, the wide range of scaling observed in the energy spectra can be compared to systems not yet fully understood theoretically, such as sperm suspensions \cite{Turbulence_of_swarming_sperm}, where the kinetic spectra exhibit similar small length decay as our model at certain self-propulsion speeds. The energy spectra also highlights the change of the systems' characteristic length scale from intermediate scales to large scales via a shift in the spectra maxima. 

We further characterise the enstrophy spectra, $\Omega_q \propto \int \langle \lvert w(q) \rvert^2 \rangle d \Vec{q}$, as the limit of $q \rightarrow 0$ in the enstrophy spectrum is directly related to the long-distance decay of vorticity correlations. The inset in Fig. \ref{fig: Flow}b shows the decay of the vorticity correlations in real space, $\langle \omega^2(r) \rangle \propto r^{-\nu}$, for large distances ($r \rightarrow \infty$). We note that the exponent $\nu$ is related to the exponent obtained from enstrophy spectra, $b$, by $\nu=b+1$ using the relation between vorticity spectrum and enstrophy, and applying Fourier relations. This vorticity scaling is very interesting and indicates that $V_0$ induces long-range vorticity order as the decay exponent at long distances $r \rightarrow \infty$ ($q \rightarrow 0$), $\nu < 2$, is smaller than the dimension of the system, $d=2$. Importantly, both inset plots exhibit a change in behaviour around $V_0=0.06$, consistent with the critical point found during the previous sections (Figs. \ref{fig: OrderEnhancment}d, \ref{fig: OrderEnhancment}e and Fig. \ref{fig: Anti-hyperuniformity}b).

\begin{figure}[htb!]  
    \centering
    \includegraphics[width=0.5\linewidth]{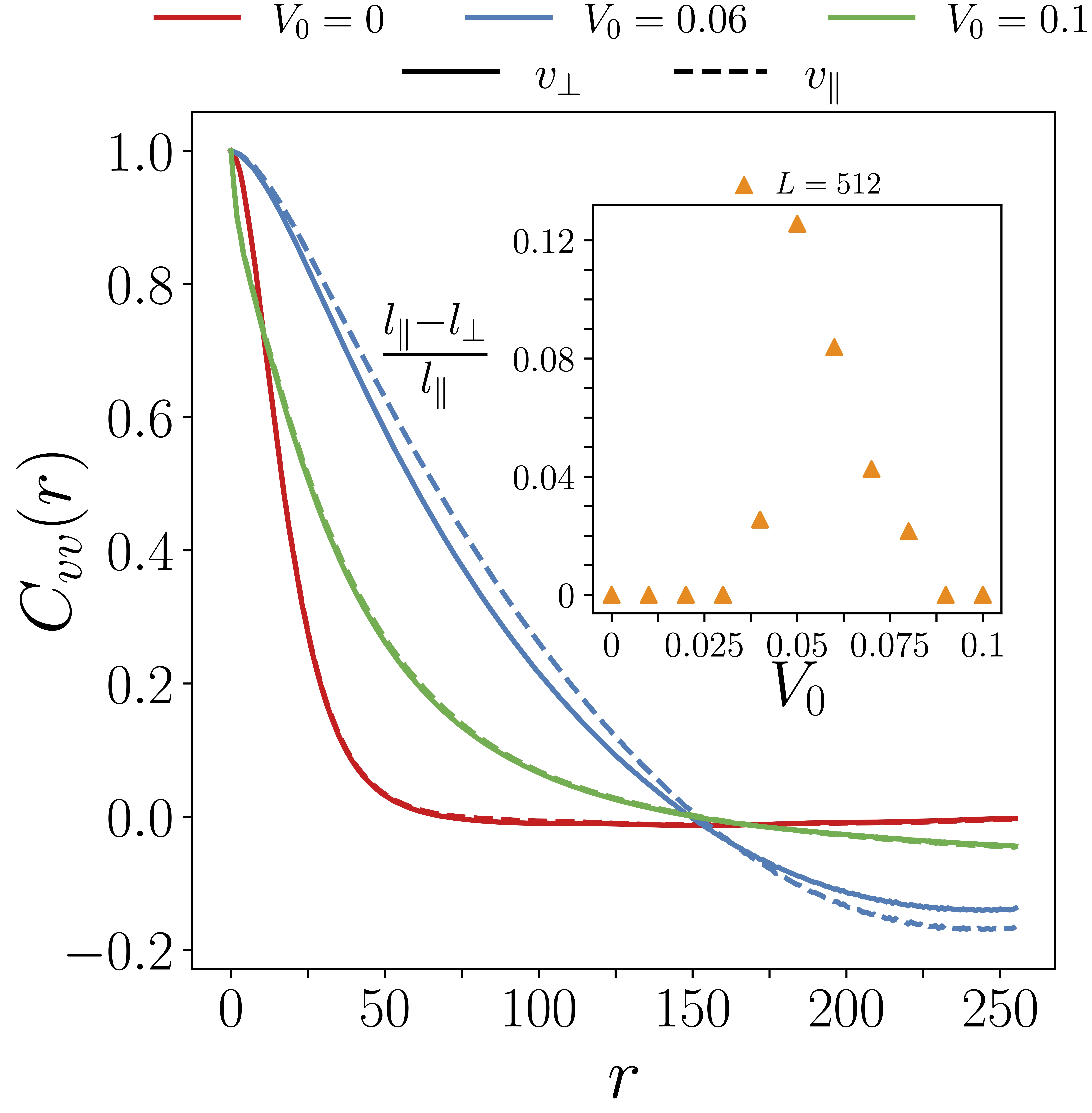}
    \caption{\textbf{Rotational symmetry breaking.} Velocity correlation function, $C_{vv}(r)$, as a function of the radial distance for different self-propulsion coefficients, $V_0$. Displayed for both the parallel, dashed line, and perpendicular velocity, continuous line, components. Averaged over different conformations. System length, $L=512$. }
\label{fig: Correlation velocity}
\end{figure}

For a purely nematic system, $V_0 = 0$, the system maintains its rotational symmetry invariance SO(2), meaning that there is no preferred direction. However, the new term $V_0 p_k $ breaks the nematic symmetry as it assigns polarity (a direction) to the nematic director $\hat{n}$. To explore how introducing self-propulsion terms influences the rotational symmetry of the system we define the perpendicular and parallel coordinate system with respect to the local direction of the flow. We rotate the velocity components to this new system of coordinates: $v_x,v_y \rightarrow v_\parallel , v_\perp$. We then find the autocorrelation function for both. We repeat and average this process for different lattice sites and conformations.
We observe that the curves for $V_0=0$ and $V_0=0.1$ (\textit{red and green lines} in Fig. \ref{fig: Correlation velocity}) overlap greatly. This suggests that the system retains rotational symmetry. Interestingly, however, for  $V_0=0.06$ (\textit{blue line} in Fig. \ref{fig: Correlation velocity}) the parallel and perpendicular correlations deviate from each other and the parallel component remains correlated over longer distances than the perpendicular component, suggesting that the rotational symmetry of the system has been spontaneously broken, and the system has become anisotropic. The inset plot of Fig. \ref{fig: Correlation velocity} shows the difference between the parallel, $l_\parallel$, and perpendicular, $l_\perp$, correlation lengths as a function of $V_0$. Both correlation lengths are similar for large and small self-propulsion coefficients, nonetheless for intermediate regimes the correlation lengths differ significantly. This indicates that self-propulsion, $V_0$, has a re-entrant symmetry-breaking behaviour. As $V_0$ increases to a first threshold value, the rotational symmetry of the system is broken. However, upon further increasing $V_0$ to a second threshold value, the symmetry is restored. The self-propulsion values that break the systems' rotational symmetry are directly linked to those that enhance long-range order, $V_0 \approx 0.06$.

\section{\label{sec: conclusion} Conclusions}
In this study, we have developed and analysed a simple model that incorporates self-propulsion into the nematic formalism, transforming individual particles from immobile shakers into self-propulsive rods. This work is inspired by the increasing number of experimental realizations of polar particles that exhibit half-integer defects and are described using the nematic formalism despite their broken symmetry \cite{PNAS_living_liquid_crystals,Blanch_Mercader_2018}. Our linear stability analyses show how self-propulsion can suppress the nematic instability \cite{Giomi_LSA_Nematic}, providing new insights into the mechanisms affecting this instability. Furthermore, we demonstrate that the self-propulsion speed can regulate the systems' order, creating long-range order in a specific range of self-propulsion speeds. The emergence of larger nematic ordered regions affects the conformation of defects, which exhibit anti-hyperuniformity and giant number fluctuations near critical values. By computing energy and enstrophy spectra, we confirm the enhancement of long-range order and the presence of non-universal scaling exponents. Additionally, we establish that the addition of a polar term, such as self-propulsion, can generate anisotropy and break rotational symmetry in a nematic system.

These results open new pathways for describing systems that share nematic symmetry and self-propel, such as cell tissue monolayers that exhibit nematic order while retaining some polar features like cell migration \cite{epithelial_govern_cell_death_2017,Benoit_Cell_migration_2021}. This framework may also be applied to modelling bacteria, such as \textit{Myxococcus xanthus}, that maintain their motile behaviour while aligning in either a nematic or polar manner \cite{han2023local} or \textit{Pseudomonas aeruginosa}, that can polarize based on the concentration gradient along their body length \cite{Chemotaxis_P_aer_2024}.

Looking forward, our model provides a framework for exploring the interplay between nematic order and self-propulsion in active matter systems. Future research could focus on experimental validation of our predictions and the exploration of other polar terms that may influence nematic systems. Bridging the gap between self-propulsive and nematic systems paves the way for designing synthetic systems with tailored collective behaviours, offering exciting possibilities for advancements in materials science and biology. 

\vskip6pt

\section*{Competing interests}
We declare no competing interest. 
\section*{Author contributions}
A.D. designed the project. N.dG.S. performed linear stability analyses, numerical simulations, and analyses of order and flow, and S.G.A. did the hyperuniformity analyses under the supervision of A.A. and A.D. All authors contributed to the writing of the manuscript.
\begin{acknowledgments}
We thank members of the Active Intelligent Matter Group for helpful comments and discussions.
\end{acknowledgments}

\section*{Funding}
This work was further supported by the Novo Nordisk Foundation grant no. NNF18SA0035142, NERD grant no. NNF21OC0068687 (AD), the Villum Fonden Grant no. 29476 (AD), and the European Union via the ERC-Starting Grant PhysCoMeT (AD). A.A. acknowledges support from the EU’s Horizon Europe research and innovation program under the Marie Sklodowska-Curie grant agreement No. 101063870 (TopCellComm).

\appendix
\section{Linear stability analysis \label{sec: LSA appendix}}

The Jacobian matrix is given by 
\begin{equation}
    \label{eq: jacobian matrix}
    \mathcal{J}_{nm}=
    \begin{bmatrix}
        A_{nm} & B_{nm} \\
        C_{nm} & D_{nm}
    \end{bmatrix},
\end{equation}
with the components having the following form:
\small
\begin{equation}
    \begin{split}
        A_{nm}=-q^2 \Gamma K (n^2+m^2) -i q V_0 n, \\
        B_{nm}=\frac{\lambda}{2}\frac{n^2-m^2}{n^2+m^2}+1, \\
        C_{nm}=-\frac{q^4 K (n^2+m^2)}{\rho}[n^2(2+\lambda)+m^2(2-\lambda)]+q^2\frac{\zeta}{\rho} (n^2-m^2), \\
        D_{nm}=-\frac{\eta}{\rho}q^2(n^2+m^2).    
    \end{split}
\end{equation}
\normalsize
The instability rises from the longitudinal mode, $(n,m)=(1,0)$. When introducing it into the Eq. \ref{eq: jacobian matrix}, one obtains the longitudinal Jacobian matrix:
\begin{equation}
    \label{eq: jacobian matrix long}
    \mathcal{J}_{10}=\begin{bmatrix}
        -q^2 \Gamma K -i q V_0 & \frac{\lambda}{2}+1 \\
        -\frac{q^4 K}{\rho}[(2+\lambda)]+q^2\frac{\zeta}{\rho} & -\frac{\eta}{\rho}q^2
    \end{bmatrix}.
\end{equation}
The eigenvalues of this matrix are given in Eq. \ref{eq: eigenvalue long}.

\section{Estimating the structure factor} \label{sec: sfac appendix}

\noindent
As is common when working with long-range density fluctuations of point configurations, the spatial distribution of topological defects at steady state is assumed to be generated by a translationally invariant and ergodic point process \cite{torquato2018hyperuniform}.
The former assumption entails that the pair correlation function, $g_2$, satisfies $g_2(\mathbf{r}_1 + \mathbf{y},\mathbf{r}_2 + \mathbf{y})= g_2(\mathbf{r}_1,\mathbf{r}_2)$ for any $\mathbf{y} \in \mathbb{R}^d$ and thus $g_2(\mathbf{r}_1,\mathbf{r}_2) = g_2(\mathbf{r}=\mathbf{r}_2 - \mathbf{r}_1)$. With this assumption, the static structure factor in the infinite volume limit is given by $S(\mathbf{q}) = 1 + \rho_I \int_{\mathbb{R}^2} d\mathbf{r} ~ (g_2(\mathbf{r}) - 1) e^{-i\mathbf{q}\cdot \mathbf{r}}$, where $\rho_I$ is the number density in the infinite volume limit \cite{torquato2018hyperuniform}.

Assuming ergodicity implies that any realization of the ensemble is representative of the ensemble in the infinite volume limit, in the sense that volume averages in this limit equal the corresponding ensemble averages \cite{torquato2018hyperuniform}. In particular, we can approximate $\rho_I$ by the ensemble average of the number density $\langle \rho_N \rangle$.

On a square domain with side length, $L$, the infinite-volume expression for $S$ can be estimated by \cite{hawat2023sfactor}
\begin{align*} 
    \hat{S} ( \mathbf{q}) = \frac{1}{\langle \rho_N \rangle L^2} \left| \sum_{j=1}^N e^{-i \mathbf{q} \cdot \mathbf{x}_j} \right|^2, ~~~~ \mathbf{q} \in \mathbb{A}_L,
       \label{eq:methods:sfac}
\end{align*}
where $\mathbf{x}_j$ refers to the point positions, $\mathbb{A}_L$ denotes the set of allowed wavenumbers, i.e. the set of $\mathbf{q}$ for which each component $q_i = 2\pi n /L$ for $n \in \left\{1,2,...\right\}$. This estimate is asymptotically unbiased, in that the ensemble average $\langle \hat{S} ( \mathbf{q}) \rangle $ converges to the true structure factor $S(\mathbf{q})$ as $L \rightarrow \infty$. 

\section{Estimating the local defect density and its moments}\label{sec: number variance appendix}

For each frame, we count the number of defects within a randomly placed spherical window centred at $\mathbf{x_0} \in [R_{\textup{max}}, L - R_{\textup{max}}] \times   [R_{\textup{max}}, L - R_{\textup{max}}]$ with radius $R_i$, and for computational efficiency, the centre point $\mathbf{x_0}$ is kept fixed while varying $R_i$ from $R_{\textup{min}}$ to $R_{\textup{max}}$. 
To avoid bias, the area of the union of all windows in a frame should be less than half that of the system \cite{torquato2021numberfluctuations}, which with just one window puts an upper limit on $R_{\textup{max}}$ as $R_{\textup{max}}=L/\sqrt{2\pi}$.
Having found the number of points within a circle with radius, $R_i$, for each frame, the average number of points and higher moments can be estimated empirically.

\vskip2pc

\bibliographystyle{RS}
\bibliography{Bibliography}

\end{document}